\documentstyle[11pt,epsf]{article}

\input psfig
\def\beq{\begin{eqnarray}}
\def\eeq{\end{eqnarray}}
\def\bea{\begin{eqnarray*}}
\def\eea{\end{eqnarray*}}

\def\centeron#1#2{{\setbox0=\hbox{#1}\setbox1=\hbox{#2}\ifdim
\wd1>\wd0\kern.5\wd1\kern-.5\wd0\fi
\copy0\kern-.5\wd0\kern-.5\wd1\copy1\ifdim\wd0>\wd1
\kern.5\wd0\kern-.5\wd1\fi}}
\def\ltap{\;\centeron{\raise.35ex\hbox{$<$}}{\lower.65ex\hbox{$\sim$}}\;}
\def\gtap{\;\centeron{\raise.35ex\hbox{$>$}}{\lower.65ex\hbox{$\sim$}}\;}

\def\singleandthirdspaced{\baselineskip=\normalbaselineskip\multiply
    \baselineskip by 130\divide\baselineskip by 100}

\newcommand{\newc}{\newcommand}
\newc{\qbar}{{\overline q}}
\newc{\Kahler}{K\"ahler }
\newc{\deltaGS}{\delta_{\rm GS}}
\begin{document}
\begin{titlepage}
\begin{flushright}
{\large hep-th/0010035 \\ SCIPP-00/29\\
}
\end{flushright}

\vskip 1.2cm

\begin{center}

{\LARGE\bf The Utility of Quantum Field Theory\footnote{Plenary Talk
on Quantum Field Theory at ICHEP 2000, Osaka, Japan.}}

\vskip 1.4cm

{\large  Michael Dine}
\\
\vskip 0.4cm
%{\it $^a$Stanford Linear Accelerator Center,
%     Stanford CA 94309} \\
{\it Santa Cruz Institute for Particle Physics,
     Santa Cruz CA 95064  } \\
%{\it $^c$Physics Department,
%     University of California,
%     Santa Cruz CA 95064  }

\vskip 4pt

\vskip 1.5cm

\begin{abstract}
This talk surveys a broad range of applications of quantum field
theory, as well as some recent developments.  The stress is on the notion
of effective field theories.  Topics include
implications of neutrino mass and a possible small value of
$\sin(2\beta)$, supersymmetric extensions of the standard model,
the use of field theory to understand fundamental issues in string
theory (the problem of multiple ground states and the question:  does string
theory predict low energy supersymmetry),
and the use of string theory to solve problems in field
theory.  Also considered are a new type of field theory,
and indications from black hole physics
and the cosmological constant problem that effective field theories
may not completely describe theories of gravity.
\end{abstract}

\end{center}

\vskip 1.0 cm

\end{titlepage}
\setcounter{footnote}{0} \setcounter{page}{2}
\setcounter{section}{0} \setcounter{subsection}{0}
\setcounter{subsubsection}{0}

%%%%%%%%%%%%%%%%%%%%%%%%%%%%%%%%%%%%%%%%%%%
%%%%%%%%%%%%%%%%%%%%%%%%%%%%
\singleandthirdspaced

%\begin{document}

\section{Introduction:  Why a Talk on Quantum Field Theory?}

Before the advent of string theory, theorists tended to
distinguish themselves as ``phenomenologists" and ``quantum field
theorists."  Phenomenologists were rather lowly sorts, who dealt
with questions having to do with experiment; the quantum field
theorists dealt with ``deep" questions such as anomalies and solitons.
Presumably this is why it is traditional at this meeting to
have sessions on quantum field theory and a summary talk.  Today,
the dividing line is more between string theory and phenomenology.
It is a rare theorist who describes him or herself as a quantum
field theorist.  There is, however, some logic to having these sessions and
this talk.  First, quantum field theory is an essential tool
both to those interested in phenomenological questions and those
interested in the difficult questions of string theory and quantum
gravity.  Second, despite its current, orphaned status, it remains
a subject of great fascination in itself, and great strides
continue to be made in its development.

This reasoning gives the speaker license
to speak about anything and everything, from phenomenology to string
theory, and that is what I will
do.  Below is an outline of the topics to be covered:
\begin{itemize}
\item
Quantum Field Theories as Effective Theories
\item
Quantum Theory and Experiment:  The Standard Model
\item
Quantum Theory and Experiment:  Beyond the Standard Model

A.  Neutrinos

B.  What if $\sin(2 \beta) \approx 0$ (more precisely, what are the
implications of a small CP asymmetry in $B \rightarrow \psi K_s$?)

\item
Applications of Quantum Field Theory to Fundamental Problems in
Physics:  String (M) Theory
\item
Applications of String Theory to Problems in Quantum Field Theory
\item
New Ideas in Quantum Field Theory:  Large Dimensions,
Non-Commutative Field Theory
\item
Limitations of Quantum Field Theory:  The Cosmological Constant
Problem.
\item{Beyond Quantum Field Theory:  Holography}
\end{itemize}

%\begin{prop}

\section{Quantum Field Theory as
Effective Theory}

In the past 30 years, quantum theory has emerged triumphant
as the framework in which to describe nature
in the very small. At the same time, we have come
to view quantum field theories as {\it effective} theories, valid
up to some energy scale.  This idea is familiar in the Fermi
theory, a theory which breaks down at scales of order
$10$'s of GeV, where it is supplanted by a larger theory.

The development of this viewpoint may account for the
current lowly status of these theories, but this is somewhat
unfair, because effective field theory accounts for virtually
everything we currently understand about nature. The standard model
follows from simply postulating a gauge symmetry and
particle content.  Renormalizability is not a principle to
be enforced, but rather the effects of non-renormalizable
operators are
suppressed by some scale (compare $250$ GeV
in the Fermi theory), $\Lambda$, where some new particles, interactions,
or other phenomena must appear.
The effects of
physics at scales above $\Lambda$ can be absorbed into the
parameters of the effective lagrangian of the low energy field
theory.

These ideas have a utility which extends beyond applications to
the standard model.  They give us a framework in which to
understand the limitations of the model, and also the physics
which may lie beyond.  They also provide powerful tools
to understand basic
questions which we confront in theoretical physics.
By thinking
about the low energy behavior of theories which, at a macroscopic
level, are very complicated,  we have been
able, during the last several years, to:

\begin{itemize}
\item
Make significant progress in understanding field theories
themselves.  It has proven possible to make exact statements about
a variety of field theories, even strongly interacting ones,
particularly in cases where the theories are supersymmetric.  There
are quantum field theories where we can study phenomena such as
confinement and electric-magnetic duality in {\it controlled}
approximations.
\item
Make exact statements about strongly interacting limits of
string theory, even though we don't have a non-perturbative setup
in which to describe the microscopic theory.
\end{itemize}

\section{  The Standard Model as an
Effective Field Theory}

Physics as we know it has $SU(3)\times SU(2) \times U(1)$
symmetry.  There are three generations of quarks and leptons.  The
unsuppressed (i.e. renormalizable) terms are exactly the gauge and
Yukawa interactions of the standard model
% (assuming, say, one
%Higgs doublet).
In addition, there are a variety of possible suppressed -- non-renormalizable
-- interactions.  The most interesting of these -- those which we
have the best chance of observing -- are those which violate
cherished symmetry principles, e.g. baryon number, such as:
\begin{equation}
{\cal L}_{\not B} ={1 \over \Lambda^2} Q Q Q L
\end{equation}
or lepton number:
\begin{equation}
{\cal L}_{\not L}={1 \over \Lambda} \phi L \phi L
%\end{prop}
\label{neutrinooperator}
\end{equation}
%\end{itemize}

In each case, the question is:  what is $\Lambda$?  If we are lucky, $\Lambda$
is not so large that we can't observe the corresponding
phenomenon.  We have some theoretical guesses:  $\Lambda$ might be the
Planck scale of the scale of grand unification. It might be some
scale associated with supersymmetry breaking ($10^{11}$ GeV?
$10^3$ GeV?)

The operator of eqn. \ref{neutrinooperator} gives rise to neutrino
masses.  The increasing evidence for neutrino masses suggests that
$\Lambda$ is not too small.  We might expect, for example, that
this operator is suppressed by quark
or lepton masses, just as are typical Yukawa couplings.
 For example, we might guess they are suppressed by
 $y^2, ~y \sim m_{\tau}/v$, so
 \beq
m_{\nu} = { m_{\tau}^2 \over \Lambda} \eeq
In this simple-minded
view, neutrino masses suggest that there is a new scale in nature,
perhaps at $10^{11} GeV$.
Of course, without a theory, $y$
can vary over a huge range, and so can $\Lambda$.  Still,
neutrino masses are our first glimpse at a new scale of
physics -- real physics beyond the standard model.

\section{Standard Model and CP Violation?}

Shortly before coming to this conference, one of my experimental
colleagues asked me what I would think if
$\sin(2 \beta) \ll 1$.  He seemed to believe that I would be very
troubled by such a finding, since it is not compatible with the
standard model.  Instead, my reaction was one of great excitement.
First,  such an observation would have a natural description in the language
of effective field theory.  Second, it would definitely mean that there is new
physics, at a not too distant energy scale.  Third, various
people, myself included, have long proposed that this is quite a
reasonable -- even likely -- possibility, if nature exhibits
low energy supersymmetry\cite{nirreview,approximatecp,othercp}.
Let me explain each of
these points.

If $\sin(2\beta)$
is small, then so is the CKM phase.  In the limit
of vanishing CKM phase, there is no CP violation in the standard
model, so there must be some new physics.  (In this situation,
of course, the measurement of the CP asymmetry in $B \rightarrow
\psi K_s$ is not necessarily a measurement of $\sin(2 \beta)$,
since there are contributions to this process beyond
those of the standard model.)  In the $K - \bar K$ and
$B - \bar B$ systems, it must be possible to describe the effects
of this new physics by operators in the low energy effective
lagrangian.  An operator such as
\beq
{\cal L}_{\Delta s =2} = {e^{i \phi} \over \Lambda^2} {s \bar d} {s \bar
d}+ {\rm h.c.}
\eeq
for example, could be responsible for $\epsilon$, where $\Lambda$,
as always, represents the scale of the new physics.

Now suppose that nature is supersymmetric, with supersymmetry
broken at a scale
$\Lambda = M_{susy} < TeV$.  In supersymmetry, there are many
new sources of CP violation.
CP-violating, $\Delta s=2$ operators are generated
in the low energy theory, for example, by exchanges of gluinos and
squarks.   The possibility that phases in the squark
and gluino mass matrices might be the origin of the observed
CP violation has been explored for some time\cite{approximatecp},
and was discussed by Ko at this meeting\cite{ko}.

Not only is it {\it possible} that this is the case, but one might
even argue that it is {\it likely}.  The line of argument goes as
follows.  If nature is supersymmetric with
supersymmetry broken near the weak scale, one typically obtains
too large a value for the neutron electric dipole moment, unless
CP-violating phases are small, of order $10^{-2}$.
CP violating phases can naturally
be small if CP is a good symmetry of the microscopic theory,
spontaneously broken at some lower scale.
In this case, the KM angle is small.
This picture
finds support in string theory, where $CP$ is a gauge
symmetry, which must
be spontaneously broken\cite{wittenstrominger,dlm}.

The final ingredient in this picture comes from the physics of
flavor conservation/violation.  In supersymmetric theories, the
absence of flavor-changing neutral currents is not automatic; some
additional structure must be imposed.  In many of the proposals
for this additional structure, $K-\bar K$ mixing is nearly
saturated by supersymmetric contributions, so the supersymmetric
contribution to $\epsilon$ is automatically of the correct order
if the CP-violating phases are of order $10^{-2}$
($\epsilon^{\prime}$ can also be accomodated).

So the answer to my colleague is:  few things could be as exciting
as the discovery that $\sin(2 \beta)$ doesn't agree with the
Standard Model prediction, and there is at least one
well-motivated framework which predicts that $\sin(2 \beta)$
should be quite small.

\section{The Hierarchy Problem}

Thinking about the Standard Model as an effective field theory
with a cutoff $\Lambda$
leads immediately to a puzzle connected with the Higgs field.
Dimensional analysis implies
\beq
m_H^2 = c \Lambda^2,
\eeq
and this is obtained from the Feynman graphs of the effective
theory.  The fact that $m_H < TeV$ (and is most likely much
lighter) suggests that new physics should not be too far away.

Previous guesses about this physics were:
\begin{itemize}
\item
New Strong Interactions -- Technicolor. Here $\Lambda \sim 1 {\rm
TeV}$.
\item
A New Symmetry -- Supersymmetry.  Here, we expect $\Lambda \sim
M_z- 1 {\rm TeV}$.
\end{itemize}

Of course, one of the likely possibilities has always been:
something we have not guessed.  In the past two years, there has
been extensive
exploration of two new possibilities:
\begin{itemize}
\item
Large extra dimensions.  This was the subject of Lawrence
Hall's excellent review at this meeting. The basic idea here
is that $m_H$ {\it is} the fundamental scale; the Planck
mass is large because some or all of the extra dimensions
are large\cite{hv,largedimensions}.
\item
Warped dimensions:  here the idea is that the hierarchy of scales results
from an exponential dependence of the metric on the distance in an
additional dimension\cite{rs}.
We will discuss this possibility further
shortly.
\end{itemize}

\section{Supersymmetry and Our Understanding of Field
Theory}

Field theories such as real QCD are complex, and it is difficult
to extract even qualitative information.
In the last few years, however, it has been possible
to solve, at least in part, many non-trivial
field theories with supersymmetry.  Supersymmetry
turns out to give a great deal of mathematical
control.  This has allowed an attack
on basic issues in quantum field theory such as duality and
confinement, and on the basic problem of supersymmetry
phenomenology: understanding the origin of supersymmetry breaking.

In supersymmetric theories, the coupling constants are often
{\it complex numbers}; the reason one can learn so much about
these theories, is that many physical quantities are
analytic functions of these numbers.

Supersymmetric gauge theories provide an example of this
phenomenon\cite{schifmanetal}.  The imaginary part of the coupling
constant is the $\theta$ parameter.  The low energy effective
coupling is an analytic function of the ``bare coupling,"
\beq
\tau = {8 \pi^2 \over g^2} + i \theta~~~~~~~g_{eff}^{-2}=
f(\tau).
\eeq

This, by itself, does not allow one to say much.   But the
the low energy theory is often symmetric under: \beq \tau
\rightarrow \tau + 2 \pi i \eeq
(corresponding to the $2 \pi$ periodicity of the $\theta$
parameter).
This highly restricts
$f$;
\beq
f = \tau +  \sum_{n>0} a_n e^{-n \tau}
\eeq
This equation, for example, says there are no corrections to the
coupling constant in perturbation theory (the
subtle meaning of this
statement was made clear in \cite{schifmanetal}, where
it was shown how one can compute a $\beta$-function
to all orders in perturbation theory).
Further considerations in some cases determine $f$
completely.

This type of analysis has given control many seemingly
impossible problems in supersymmetric field theories.
These include:
\begin{itemize}
\item Exact solutions of theories with N=2 supersymmetry
(Seiberg and Witten\cite{sw}).  In these theories many
quantities can be computed exactly.  In the strongly coupled
region, for example, one can study confinement as
the result of monopole condensation.  Further
progress in this area was reported at the meeting, including
exquisite tests of these ideas (Fucito\cite{fucito}, Khoze\cite{khoze})
and determination
of patterns of symmetry breaking in particular
examples (by Murayama\cite{murayama} and Yasue\cite{yasue}).
\item Theories with N=1 Supersymmetry might provide the
solution to the hierarchy problem.  Not only do they give a way
of understanding why there are not big corrections to the Higgs
mass $\Lambda \ll M_p$, but they can naturally
produce very large hierarchies.  Further progress on such theories
was reported at this
conference (Kazakov\cite{kazakov}, Nitta\cite{nitta},
Tachibana\cite{tachibana}).
\end{itemize}

\section{Applied Duality}

Apart from addressing fundamental questions in field theory, we
can try to use these ideas to understand, e.g., how supersymmetry
might be realized in nature.  This is an area which has been
developing for some time, but the past two years have seen some
interesting new ideas:

\begin{itemize}
\item
Models with dynamical supersymmetry breaking have been
put forward in which the partners of the first two generations of fermions
are composite and quite massive, while the partners of the third
generation are light\cite{composite}. These models,
first, implement both the ideas of dynamical supersymmetry
breaking and quark and lepton compositeness.  They are readily
compatible with bounds
coming from direct searches as well as
processes such as $b \rightarrow s + \gamma$.
They are simpler and less contrived than earlier proposals.
\item Models in which nature is
approximately conformally invariant over a range of energies can
address not only
supersymmetry breaking, but also provide models of flavor\cite{ns}.
Yukawa hierarchies arise because
fields in different generations possess different anomalous
dimensions.  Many problems of flavor physics are readily
understood in this context.
\end{itemize}

\section{String Theory as a Tool for the
Investigation of Field Theory}

Over the past several years, it has proven fruitful to consider
certain problems in quantum field theory from the perspective of
string theory. This is illustrated by simple configurations of
$D$-branes.
In Type II string theory, a configuration of $N$ parallel
$D3$ branes describes a theory
with gauge group $SU(N)$ and $N=4$ supersymmetry.  More
interesting models, with less supersymmetry, can be constructed
along these lines.  Problems which are very difficult from a field
theory perspective take on quite a different character (e.g.
geometric) in the string picture.

Recent new developments have been based on the
``AdS-CFT" correspondence.  This correspondence asserts that
conformally invariant QFT is equivalent to string theory
in AdS space\cite{adsreview}.
This can be used to provide insight into a variety of field
theories with conformal invariance, but
we would like to understand real QCD, which is certainly not
conformally invariant.  It is necessary to perturb the system in
some way.

Early approaches to this problem involved,
for example,  finite temperature in five dimensions (in the high
temperature limit, the system becomes essentially four
dimensional)\cite{adsreview}.  These methods were of limited
power.
Recently, Polchinski and Strassler have exhibited cases
where one can perturb the conformal theory by adding
non-conformally invariant operators, and where the physics
on the supergravity side is completely under control (non-singular
spaces)\cite{polchinskistrassler}.  These are
theories where confinement, flux tubes, glueballs, and other interesting
phenomena can be thoroughly studied in the gravity dual.

\section{A New Type Of Field Theory}

In the past year, much attention has been focused on a new
type of field theory, known as ``non-commutative field theory" (NCFT).
These theories arise in some cases as the low energy limits of
string theories, and seem to incorporate some of the non-locality
of string theory\cite{noncommutative}.
They exhibit
bizarre connections between the infrared and the ultraviolet.
These features are interesting in themselves, and might
be relevant to understanding difficult problems such as the
cosmological problem and issues in black hole physics.

The basic feature of these theories is that space coordinates do
not commute: \beq [x,y] = i \theta. \eeq
This sort of
relation arises in string theory in the presence of a background
magnetic field. NCFT's can't be local.  They exhibit
peculiar connections between the infrared and ultraviolet -- which
have come to be called the {\it infrared-ultraviolet}
connection.  For example,
typical Feynman graphs behave as \beq \int {d^4 k \over (2 \pi)^4}
{1 \over k^2} e^{i\theta  p_1 k_2}
\eeq
For $\theta=0$, this diagram would be highly divergent in the
ultraviolet, but for $\theta \ne 0$, it behaves as
$ {\theta \over p^2}$.  In other words,
an ultraviolet divergence gets replaced by a divergence
as $p^2 \rightarrow 0$.

It is fair to say that the significance of these
theories is only beginning to be understood. Could there be
real phenomena which might be described by such theories?  Might they give
some insight into the cosmological constant problem?  Could these
structures have relevance to other areas of physics?  Time will
tell.%(Kuirke\cite{kuirke} and Chaichian\cite{chaichian}
%reported results on the renormalization of these
%theories.

\section{Field Theory as a Tool for
Understanding String
Theory}

The pictures which have been described above can be viewed
from a different perspective:  One can hope to use one's
understanding of field theory in order to understand difficult
questions in string (M) theory. (See the talk of Paul
Townsend at this meeting.)

These ideas have a long history.  The easiest way to prove the
finiteness of string theory is to study the effective field
theory.  Indeed, even though there is much that we do not
understand about the fundamental structure of the theory, many
questions can be addressed by considering the low energy field
theory limit.

Here are just a few of the areas in which field theory has proven
useful to understanding outstanding problems in string theory:
\begin{itemize}
\item
Much of the understanding of duality in string theory has
been obtained from the study of the low energy effective field
theory.
\item
String theory has a host of possible vacuum states which
are uncovered in various approximations.  These are characterized
by the number of dimensions (2-11), the amount of
supersymmetry $(N=0, \dots 4)$, the number of
generations, as well as sets of continuous parameters
(``moduli").  The hope is that some dynamical effects
pick out one vacuum or another.  {}From
considerations of the low energy effective field theory, however,
we know that all of the vacua with some supersymmetry in $d \ge
5$ or with $N>1$ supersymmetry in $d=4$ are true, stable vacua of string
theory, {\it exactly}.
\item  We can make many exact statements about more promising
vacua which, in some approximation,
have $N=1$ supersymmetry.  We can often
compute the ground state energy as a function
of the moduli reliably using effective
field theory. We can sometimes argue that
couplings unify even if the theory is strongly coupled.

\end{itemize}

\section{Unconventional Approaches to
Outstanding
Problems}

\subsection{Large Dimensions}

In the past two years, two new approaches have been put forth
to the hierarchy problem.  While the underlying justification for
both is string or M theory, both are firmly based on pictures
developed by considering the low energy field theory.

The premise of each of these proposals is that the fundamental scale of
physics might be close to the weak scale. This obviates the need
for supersymmetry as a solution to the hierarchy problem, and,
indeed, in both of these approaches, low energy supersymmetry (at
least as it is conventionally discussed) is not a likely outcome.

Lawrence Hall has discussed the large dimension possibility at
some length at this meeting.  The basic idea is that the fundamental
scale of the theory is of order a TeV.  The Planck scale, in this
view, is large because some set of extra dimensions are large.
I will not review this proposal in detail here.
However, I would like to mention two sets of ideas about
supersymmetry breaking which have emerged from thinking about
large, but not extremely large, extra dimensions.  These start from the
idea of two separated walls, with the standard model on one wall,
supersymmetry-breaking on another.
The first of these is known as Anomaly Mediation\cite{anomalymediation}.  Precursors
of
this idea arose from four-dimensional, field theoretic
reasoning\cite{dinemacintire}.  In this picture, one finds an
approximate degeneracy between squarks, necessary to understand
the suppression of flavor violating processes.  In the simplest
version, some sleptons are tachyonic, however, and it is necessary
to consider rather complicated models.
The second is known as gaugino mediation\cite{gauginomediation}.
Again, this idea has field theoretic precursors\cite{dks}, but finds a
firmer motivation in the large dimension picture.  Here, the idea
is that certain gauge multiplets propagate in the bulk, and are
natural candidates to mediate supersymmetry breaking.  Again,
this is a way to obtain a spectrum with a suitable degree of
degeneracy and other distinct predictions for the low energy soft
breakings.

\subsection{Warped extra dimensions:
The Randall-Sundrum
Model(s)}

The second of these new proposals to understand the hierarchy
problem is known as the Randall Sundrum model.  Actually, there
are several versions of this model.  The simplest to describe
is set in five dimensions, with two
walls.
With the walls as sources of stress-energy, if one tunes
parameters, Einstein's equations admit a solution:
\beq ds^2 =  e^{-2 k r_o \vert  y \vert } dx^{\mu} dx_{\mu} + r_o^2 dy^2 \eeq
Four dimensions are flat, but the fifth, described by $y$, is
curved, or ``warped."  The standard model sits on the wall at
$y=1$; the wall at $y=0$ is referred to as the ``Planck Brane."

In the effective theory in four dimensions, Newton's constant is given by:
\beq G_N = {k \over M^3}{1 \over 1-e^{-k r_o \pi}} \eeq
while the typical scales on our brane are of order
\beq
m_H^2 =  M^2 e^{-2 k r_o}.
\eeq

So the hierarchy is due to the warping of space, and it is large
because it is the exponential of a rather modest
number (compare technicolor, susy approaches).
(New solutions of this type were reported at this meeting
by Ichinose\cite{ichinose}.)

What fixes the separation of the walls which determines the
exponential? Goldberger and Wise have shown that it can arise from
plausible scalar field dynamics in the low energy theory\cite{gw}.

There are a number of versions of these ideas currently being
explored.  These include the possibility that the extra
dimensions are in infinite, with gravity
localized on a brane\cite{infinitedimensions}, or that, viewed from
far enough away, the extra dimension is simply
flat\cite{gregorievrubakov}.
Surprisingly, these ideas are
not easily ruled out, and
if correct, these lead to distinctive phenomenologies (reviewed
by Hewett at this meeting\cite{hewett}),
with some features in common with the large dimension picture.

It should be noted that this structure, unlike the large
dimensions structure, has not been derived from string theory,
though there is much effort along these lines.

\subsection{An Effective Field Theory Critique}

The large dimension and warped dimension ideas are exciting, and
are plausible alternatives to supersymmetry as solutions to the
hierarchy problem.  Experiment might produce a smoking gun
for one of them.

On the theoretical side, there are many questions which must
be settled.  All of these are problems of the effective field
theory:

\begin{itemize}
\item
Proton decay, $\mu \rightarrow e + \gamma$, etc.  One can
certainly imagine various ways of suppressing proton decay.  Refs.
\cite{largeproton} provide several proposals, and
if these are operative, they provide
more than adequate suppression.  One can debate whether these are
more or less plausible than R-parity, for example, in
supersymmetric models.
\item
Flavor changing neutral currents.
\item
High precision electroweak experiments.
\end{itemize}

In each case, one expects operators to appear in the effective
low energy theory which contribute at a dangerous level, unless
the fundamental scale is sufficiently large.  Precision
electroweak experiments provided the most model-independent
limits
on the fundamental scale in the
TeV range.  This is perhaps troubling for hierarchies, and is
reminiscent of some of the problems of
technicolor.  Scenarios
have been proposed to suppress other effects; these are typically
tied in to ideas of how the KM matrix, with its various
peculiar features, is generated.  One possibility is that there is a large
flavor symmetry, with symmetry breaking occurring on branes located
far from the brane on which the standard model sits\cite{dimopoulosflavor}.
While the original models of this sort
were rather elaborate, more elegant
models were proposed in \cite{halletal}.

\subsection{Solutions to the Hierarchy Problem:  A Scorecard}

It is interesting to compare the various solutions which have been
proposed for the hierarchy problem, and to compare with the
minimal standard model.  One can score them according to:
\begin{itemize}
\item
Do they solve the hierarchy problem?
\item
Do explicit models exist?
\item
Do they explain unification of couplings in a robust, generic
way?
\item
Can they explain the absence of flavor changing processes in a
simple way?
\item
Do they explain the absence of proton decay in a simple way?
\item
Do they lead naturally to a dark matter candidate?
\end{itemize}

I will let you do the scoring yourself (I offered my own at the
conference) but I think it is clear that the standard model and
supersymmetry score the highest in any such ranking.  Still,
nature will ultimately decide.
It is hard to imagine anything which would be more exciting than
the experimental discovery of extra dimensions; I'll let you
choose where to place your bets.

\section{Is Low Energy Supersymmetry A Prediction of String Theory?}

It is often said that low energy supersymmetry is a prediction of
string theory, and indeed string theory
rather naturally produces this sort of
structure.  But the large and warped dimension ideas are plausible
alternatives and need
not exhibit the states (squarks, sleptons, neutralinos)
expected there.

{}From studies of low energy field theory limit of strings,
however, there {\it is} some evidence that non-supersymmetric states have
problems. Fabinger and Horava\cite{fh} have shown that many
non supersymmetric states of string theory undergo catastrophic
decay.  This instability is closely related to an instability of
the simplest Kaluza-Klein theory, discussed some years ago
by Witten\cite{wittenkk}.
If this problem is generic, low energy supersymmetry is a prediction of string
theory.

Are there other problems with  non-supersymmetric theories?
Perhaps related to the instability discussed above,
non-supersymmetric string theories rather typically have
tachyons somewhere in their classical moduli spaces.  One might
imagine that this is not so serious; perhaps there is simply
a nearby vacuum.  But a little thought shows that the problem is
deeper.  Even if the potential has a minimum as a function of the
tachyon field, the energy associated with this minimum is of order
$V_o = -{1 \over g^2}$.  Since $g^2$ is dynamical in M theory, the system
can attain arbitrarily low energy by moving to small enough
coupling.

All of this sounds serious, but with the current state of our
knowledge, it is hardly a proof that non-supersymmetric string
states don't make sense.  We simply don't understand string theory
well enough to decide whether it might be possible that the
universe sits in a state far from one of the tachyonic states,
or that the lifetime of the universe for the catastrophic vacuum
decay of Fabinger and Horava is much greater than the age of the
universe.  It would be interesting to exhibit some sort of disease
of the non-supersymmetric vacua, such as an anomaly, which would
decisively indicate such an inconsistency.  I have spoken about
some possible candidates for such anomalies elsewhere, and am
currently engaged in a search for examples.

%\subsection{Scorecard}
%
%This is a natural point to try and produce a scorecard, comparing
%various proposals for solving the hierarchy problem.  These
%include technicolor, supersymmetry, large dimensions, and the
%Randall Sundrum proposal.  In the talk I provided such a
%scorecard, but here I will let the reader construct his or her
%own.

\section{Limitations of Effective Field Theory?}

\subsection{Two Problems for Effective Field Theory}

There is growing evidence that the ideas of effective field theory
do not apply to gravity.  This evidence arises from the study of
Black Holes and the problem of the Cosmological Constant.

One of the most exciting
recent developments in physics is the observation of what appears to be a non-vanishing
cosmological constant, $\lambda$.  This is a quantity one would think
one could compute from particle physics.  However,
the same sort of dimensional
analysis we used before suggests that
\beq \lambda = a \Lambda^4 \eeq
So even if $\Lambda$ is as small as $100$ GeV, we obtain an
estimate $55$ orders of magnitude larger than the reported
observation!  (Alternatively, if $\lambda$ were this large, our
horizon would be about $10$ cm!)

In field theory, even if, for some reason, there is no
cosmological constant at the classical level, one expects a large
value for $\lambda$ quantum mechanically.  This is simply
because (for weak coupling) one can think of a
quantum field theory,
as a collection of harmonic oscillators, one
for each particle type and momentum $\vec k$.
The vacuum energy, which is just the cosmological constant,
then gets a contribution from the zero point fluctuations of each
oscillator:
 \beq E_o= \lambda = \sum \int^{\Lambda} {d^3 k \over (2
\pi)^3} {1 \over 2} (-1)^F \sqrt{k^2 + m^2}
\label{cosmocon}
\eeq
$$ ~~~~~~~ \propto \Lambda^4.$$
($(-1)^F$ is $+1$ for fermions, $-1$ for bosons; it arises
because, in the case of fermions, rather than considering the
zero point energy, one must compute the energy of the filled
fermi sea).
Supersymmetry might act as some sort of cutoff.  If susy were
exact, the bosonic and fermionic contributions to this expression
would cancel.  However, from our failure to observe any
supersymmetric particles to date, we know that we can safely
take the cutoff to be as large as $100$ GeV.  So the low energy
contribution to the cosmological constant is at least $56$ orders
of magnitude too large!  At our present level of understanding, we
must somehow imagine that this is miraculously cancelled (to a
part in $10^{56}$!) by high energy contributions.

Many
attempts to solve this
problem have failed.  It seems likely that this represents
a breakdown of our ideas about effective field theory.

There is a good deal of evidence that the usual rules
of quantum mechanics break down near black holes.  The problem
is connected with Hawking radiation.  Hawking showed many
years ago that black holes evaporate.  If one imagines
a black hole created in a pure state, than in the far future,
one has a thermal system.  One might imagine that this is no different
than, say, burning a piece of coal:  the original information
in the quantum system must be encoded in subtle correlations among
the outgoing photons.  This, however, turns out to violate
our usual notions of {\it locality}.
 String theory
seems to possess some degree of non-locality, and there is growing
evidence that string theory provides a consistent quantum
mechanical framework in which to understand black holes.

\subsection{  The Holographic Principle}

{}From considerations of black hole physics,
't Hooft and Susskind have suggested that in a theory with
gravity, there are not as many degrees of freedom in a volume $V$
as we might expect; they argue that a consistent theory of gravity
must be {\it holographic} -- the number of degrees of freedom is
proportional to the surface area of $V$\cite{thooft,susskind}.

This holographic principle is in many ways mysterious, but it can
sometimes be seen to hold in string theory.  At low
energies, for many purposes, string theory is well described by an
effective field theory, but perhaps not for everything?

If these ideas are correct, some important questions in
nature cannot be answered by the methods of effective field
theory.  This might be crucial to understanding not only black
holes but also the cosmological constant problem, since it means
that there are far less degrees of freedom than in eqn.
\ref{cosmocon}.
By itself, however, the holographic principle does not answer the question
of the cosmological constant.  Apart from conceptual issues, there
is a numerical one.  Even if the cosmological constant is
suppressed from some naive estimate, say $10^8 {\rm GeV}^4$, by a
factor of the current horizon in Planck units, we still miss the
observed value by more than $10$ orders of magnitude!

\section{New Ideas About the Cosmological Constant}

Can field theory in four dimensions resolve the cosmological
constant problem?

Through the years, a number of ideas have been put forward:
\begin{itemize}
\item
Perhaps the dynamics of a light particle cancels the cosmological constant,
in a manner reminiscent of the axion solution of the strong CP problem?
There have been many attempts along these lines,
but Weinberg has proven a no go theorems which shows that this cannot
occur, at least in conventional field
theories\cite{weinberglambda}.
\item
Interesting gravitational dynamics such as Euclidean
wormholes\cite{wormholes} have been proposed, but are not completely
satisfactory.
\end{itemize}

In the last few years, a number of new ideas have been put
forward.
\begin{itemize}
\item
Kachru and Silverstein\cite{kslambda}, motivated by the AdS/CFT correspondence,
have exhibited a number of string models without supersymmetry
in which the cosmological constant cancels at low orders of
perturbation theory, and have argued that this cancellation
may be general.
The known examples, however, have Fermi-Bose degeneracy and it
is not clear whether or not this is an essential part
of the proposal.
\item
As noted above, the notion of holography is likely to have
some implications for the cosmological constant problem, but
it is not clear, at present, precisely what those implications
might be.  Moreover, while the present horizon
is very large, it is not large
enough: \beq d \approx {10^{10}} ~{\rm light~ years} \eeq
so $d \times (100 {\rm GeV}) \sim 10^{36}$
(we need about
$10^{55}$).   Still, there have been some interesting suggestions
about how this might work\cite{ckn}.
\item
Warped geometries:  there have been a number of suggestions
that the gravitational equations in this framework
permit solutions with vanishing four dimensional cosmological
constant, going back to\cite{rubakovshaposhnikov}, and more
recently \cite{rslambda}.
 But troubling singularities appear, and it is not yet
clear whether these solutions make sense.  It has been argued that
these singularities are at best just a rephrasing of the fine-tuning
problems\cite{lambdacritics}.  This problem is under
intense investigation.
\end{itemize}

\section{More Embarassing Proposals}

\subsection{What are we trying to explain?}

If recent observations of a cosmological constant are
correct, then the value of the constant is just such that the
cosmological constant is becoming important in the present epoch
of cosmic history,
\beq
\Omega_{\lambda} \approx 0.7~ \Omega_{crit}
\eeq

In thinking about $\lambda$,
there is a piece of numerology about the cosmological
constant which is often invoked: \beq
\lambda \approx {({\rm TeV})^8 \over M^4}. \eeq Here $M$
is the reduced Planck mass, $M \approx 10^{18}$.

So
if we had a theory in which this relation held,
we would have $\Omega_{\lambda}$ in the right
ballpark.
But while the order of magnitude is correct,
we are confronted today with a very
close coincidence.  If we
change TeV to $2.7$ TeV, for example, in this formula
\beq
\Omega_{\Lambda} \approx 10^3 \Omega_{crit}
\label{naiveestimate}
\eeq
So even if we had a theory in which \ref{naiveestimate} held,
we would still be confronted with a significant puzzle.

\subsection{The Anthropic Principle Rears Its Ugly Head}

These remarks are suggestive of an anthropic explanation of the
cosmological constant.
I believe that the only scientifically defensible form of
anthropic explanation is what Weinberg calls the
``Weak Anthropic Principle."
Suppose a theory has many metastable (or stable) ground states.
The universe in its history may sample all of these states.
Only some may develop in a way which can allow for even
rudimentary forms of life; most might collapse, for example, long
before structure can form.

Even within this framework, as we will see, we are
treading on dangerous ground.  As Weinberg has remarked:
A physicist talking about the anthropic principle runs the
same risk as a cleric talking about pornography:  no matter how
much you say you're against it, some people will think you're a
little too interested.

How would we apply this sort of weak anthropic explanation to the
cosmological constant?  For this to make sense, the
underlying theory must have lots and lots ( $10^{120} =$ zillions and
zillions -- to borrow
a phrase from Carl Sagan) of reasonably stable ground states.  We live in one
with a small cosmological constant because that's the only place
intelligent beings can evolve.

Weinberg originally argued that this sort of weak anthropic
explanation
was not good enough to explain the cosmological constant;
this could only explain why
$\Omega_{\Lambda} < 10^2-10^3 \Omega_{crit}$\cite{weinberglambda}.
Garriga and Vilenkin
have argued that a more refined argument gives the right order
of magnitude\cite{villenkin}.
So we have to face the possibility that this {\it might} provide
an explanation of the observed facts.

Whether you like this sort of explanation or not, we need to ask:
do we know of any theories with these properties?  The short
answer is no, but recently Bousso and Polchinski,
and Donoghue have pointed
out a way in which such a vast set of metastable states might
arise in string theory\cite{bp,donoghue}.
The  analysis
is based on considerations of effective field theory, and in particular
of certain gauge fields with three indices,
$A^{[i]}_{\mu \nu \rho}$, whose flux is quantized (compare
monopoles):
\beq
F_{\mu \nu \rho \sigma}^{[i]} = q^{[i]} n^{[i]}
\epsilon_{\mu \nu \rho \sigma}.
\eeq
The vacuum energy takes on values:
\beq
E= \sum_i^{N} n^{[i]~2} q^{[i]} - \lambda_o
\eeq
where $\lambda_o$ represents the other contributions to the
cosmological constant.  The number of states grows rapidly
with $N$;
Bousso and Polchinski argue
that if $N \sim 120$, for example, there may a sufficient number of
states.

Whether these states actually exist as (meta)stable states
is an open question, but within our current understanding,
we must acknowledge that it is conceivable.  When one
delves further into this type of picture\cite{bdbp}, one finds that in
some versions, everything in this suggestion becomes anthropic.
In others, it is only the cosmological constant.
Determining whether such a vast set of states
truly exists is a problem which cannot be settled in
effective field theory.

\subsection{A Much Milder Use of the Anthropic Principle?}

Whether or not string theory has the vast set of metastable
states required for the application of the anthropic principle,
it is certain that it contains a large number of ground states,
only a small fraction of which -- if any -- resemble the real world.  We have
already noted that string theory definitely contains states with
more than four dimensions and more than four supersymmetries.
A milder application of the anthropic principle might be to
understand how nature selects among these possible vacua.
It could be that in most of them, one can not develop even
the most minimal structures one would imagine are necessary to
sustain life, and in fact that many of them would be subject to
gravitational collapse.
We are a long way from being able to answer this question
completely, but partial (positive) answers can already be given,
using methods of effective field theory\cite{bdbp}.

\section{Conclusions}

Field theory continues to enjoy an extraordinary level of
utility.  It gives the standard model, and suggests possible
extensions and new phenomena.
It gives a way of organizing our questions about new experimental
discoveries, and suggests possible explanations.
It suggests broad ranges of new phenomena.  It is a crucial tool
in our study of candidates for a fundamental theory.

Yet field theory also has limitations.  We probably need
to go beyond quantum field theory if we are to understand:

\begin{itemize}
\item
The problems of Black Holes
\item
The Cosmological Constant Problem
\item
The principles which determine the ground states of $M$ theory,
and what selects among them.
\end{itemize}

\noindent
{\bf Acknowledgements:}

\noindent
This work supported in part by the U.S.
Department of Energy.  M.D. wishes to thank Nima
Arkani-Hamed,  Jon
Bagger, Yossi Nir and Scott Thomas for
discussions and comments on the manuscript.

\end{document}